\documentclass[12pt]{article}
\usepackage{amsmath,amsfonts}
\usepackage[numbers,sort&compress]{natbib}
\usepackage{times}
\usepackage[left=2.54cm,top=2.54cm,right=2.54cm,bottom=2.54cm,bindingoffset=0.0cm]{geometry}
\usepackage{setspace}
\usepackage{mdframed}

\usepackage{graphicx}
\usepackage{amssymb}
\usepackage{vaucanson-g}

\usepackage{hyperref}

\usepackage{graphicx}
\usepackage{amssymb}
\usepackage{epigraph}
\usepackage{tikz}

\usepackage{hyperref}

\begin{document}
\parskip 0pt

\title{Wrong side of the tracks: \\ Big Data and Protected Categories}
\author{Simon DeDeo\footnote{Department of Informatics, School of Informatics and Computing, and Program in Cognitive Science, Indiana University, Bloomington, IN 47408 \& Santa Fe Institute, 1399 Hyde Park Road, Santa Fe, NM 87501. {\tt simon@santafe.edu}; {\tt http://santafe.edu/\%7Esimon}}}
\date{\today}
\maketitle

\begin{abstract}
\noindent
When we use machine learning for public policy, we find that many useful variables are associated with others on which it would be ethically problematic to base decisions. This problem becomes particularly acute in the Big Data era, when predictions are often made in the absence of strong theories for  underlying causal mechanisms. We describe the dangers to democratic decision-making when high-performance algorithms fail to provide an explicit account of causation. We then demonstrate how information theory allows us to degrade predictions so that they decorrelate from protected variables with minimal loss of accuracy. Enforcing total decorrelation is at best a near-term solution, however. The role of causal argument in ethical debate urges the development of new, interpretable machine-learning algorithms that reference causal mechanisms.
\end{abstract}
\epigraph{``SIR,---I have read your letter with interest; and, judging from your description of yourself as a working-man, I venture to think that you will have a much better chance of success in life by remaining in your own sphere and sticking to your trade than by adopting any other course.''}{{\it Jude the Obscure}, Thomas Hardy (1895)}

Should juries reason well? Should doctors? Should our leaders? When the human mind is augmented by machine learning, the question is more subtle than it appears. It seems obvious that, if we gather more information and use proven methods to analyze it, we will make better decisions. The protagonist on television turns to a computer to enhance an image, find a terrorist, or track down the source of an epidemic. In the popular imagination, computers help, and the better the computer and the more clever the programmer, the better the help.

The real story, however, is more surprising. We naturally assume that the algorithms that model our behavior will do so in a way that avoids human bias. Yet as we will show, computer analysis can lead us to make decisions that without our knowledge, judge people on the basis of their race, sex, or class. In relying on machines, we can exacerbate preexisting inequalities and even create new ones. The problem is made worse, not better, in the big data era---despite and even because of the fact that our algorithms have access to increasingly relevant information. The challenge is a broad one: we face it as ordinary people on juries, experts sought out for our specialized knowledge, and decision makers in positions of power.

To demonstrate this challenge, we will make explicit how we judge the ethics of decision-making in a political context. Through a series of examples, we will show how a crucial feature of those judgments involves reasoning about cause. 

Causal reasoning is at the heart of how we justify the ways in which we both reward individuals and hold them responsible. Our most successful algorithms, though, do not produce causal accounts. They may not reveal, say, which features of a mortgage applicant's file combined together to lead a bank's algorithm to offer or deny a loan. Nor do they provide an account of how the individual in question might have come to have those properties, or why those features and not others were chosen to begin with. The ethics of our policies become opaque. In situations such as these, decision makers may unknowingly violate their core beliefs when they follow or are influenced by a machine recommendation.

In some cases, the solution to a moral problem that technology creates is more technology. We will illustrate how, in the near-term, a brute force solution is possible, and present its optimal mathematical form. More promisingly yet, future extensions of work just getting underway in the computer sciences may make it possible to reverse-engineer the implicit morals of an algorithm, allowing for more efficient use of the data we have on hand. We describe two recent advances---contribution propagation and the Bayesian list machine---that may help this goal.

All these methods have their limits. The ethical use of machines may lead to new (short-term) inefficiencies. We may find that more mortgages are not repaid, more scholarships are wasted, and more crimes are not detected. This is a trade-off familiar to democratic societies, whose judicial systems, for example, thrive under a presumption of innocence that may let the guilty go free. Justice and flourishing, even in the presence of inefficiencies, are not necessarily incompatible. To trade short-term gain for more sacred values has, in certain historical periods and for reasons still poorly understood, led to long-term prosperity.

Our discussion will involve questions posed by modern, developed, and diverse societies regarding equality of both opportunity and outcome. We do not take a position on the merits of any particular political program or approach. Rather, our goal is to show how the use of new algorithms can interfere with the ways in which citizens and politicians historically have debated these questions in democratic societies. If we do not understand how machines change the nature of decision-making, we will find ourselves increasingly unable to have and resolve ethical questions in the public sphere.

\section{Correlation, Discrimination, and the Ethics of Decision-Making}

There are many constraints on the ethics of decision-making in a social context. We begin here with one of the clearest of the modern era: the notion of a protected category. Decisions made on the basis of such categories are considered potentially problematic and have been the focus of debate for over half a century. In the United States, for instance, the Civil Rights Act of 1964 prohibits discrimination on the basis of race, color, religion, sex, and national origin, while Title IX of the Education Amendments of 1972 makes it legally unacceptable to use sex as a criterion in providing educational opportunities.

What appears to be a bright-line rule, however, is anything but. In any society, protected and unprotected categories are strongly correlated. Some are obvious: if I am female, I am more likely to be under five feet tall. Others are less so: depending on my nation of origin, and thus my cultural background, I may shop at particular stores, have distinctive patterns of electricity use, marry young, or be at higher risk for diabetes.

Correlations are everywhere, and the euphemistic North American idiom ``wrong side of the tracks'' gains its meaning from them. Being north or south of a town's railroad line is an innocent category, but correlates with properties that a society may consider an improper basis for decision-making. What a person considers ``wrong'' about the wrong side of the tracks is not geography but rather the kind of person who lives there.

In the big data era, euphemisms multiply. Consider, for example, a health care system with the admirable goal of allocating scarce transplant organs to those recipients most likely to benefit. As electronic records, and methods for collecting and analyzing them, become increasingly sophisticated, we may find statistical evidence that properties---diet, smoking, or exercise---correlated with a particular ethnic group give members a lower survival rate. If we wish to maximize the number of person-years of life saved, should we make decisions that end up preferring recipients of a different race?

Such problems generically arise when machine learning is used to select members of a population to receive a benefit or suffer harm. Organ donation is only one instance of what we expect to be a wide portfolio of uses with inherent moral risks. Should we wish to allocate scholarship funding to those students most likely to graduate from college, we may find that including a student's zip code, physical fitness, or vocabulary increases the predictive power of our algorithms.\footnote{Note that use of physical fitness does not mean we select students who are necessarily more fit instead of more likely to graduate; rather, we can improve our original selection goal---graduation rate---by use of subtle signals that include physical fitness.} None of these properties are protected categories, but their use in machine learning will naturally lead to group-dependent outcomes as everything from place of residence to medical care to vocabulary learned in childhood may correlate with a property such as race.

In the case of the allocation of scholarship funds, we may want to exclude some sources of data as being {\it prima facie} discriminatory, such as zip code, even when they do not directly correspond to protected categories. Others, though, such as physical fitness or vocabulary, may plausibly signal future success tracking, say, character traits such as grit~\cite{duckworth2007grit,eskreis2014grit}. 

Yet the predictive power of physical fitness or vocabulary may be driven in part by how race or socioeconomic status correlates with both these predictor variables and future success. Data-mining might uncover a  relationship between adolescent fitness and future employment success. But this correlation may be induced by underlying mechanisms such as access to playgrounds that we may consider problematic influencers of decision-making. An effort to use physical fitness as a signal of mental discipline may lead us to prefer wealthier students solely because physical fitness signals access to a playground, access to a playground signals high socioeconomic status, and high socioeconomic status leads to greater access to job networks. 

Put informally, a machine may discover that squash players or competitive rowers have unusual success in the banking profession. But such a correlation may be driven by how exposure to these sports correlates with membership in sociocultural groups that have traditionally dominated the profession---not some occult relationship between a squash serve and the ability to assess the value of a trade agreement.

\section{Reasoning about Causes}

One solution to the problems of the previous section is to consider all measurable properties guilty until proven innocent. In this case, we base decision-making only on those properties for which a detailed causal mechanism is known, and known to be ethically neutral. A focus on the causal mechanisms that play a role in decision-making is well known; to reason morally, particularly in the public sphere, is to invoke causation.\footnote{See, for example, the essay~\cite{moore2010causation} or the review~\cite{sep-causation-law}, and references therein, for the role of causality in legal reasoning; more broadly, a key reason for the analysis of causation in general is its role in ethical concepts such as responsibility~\cite{sep-causation-metaphysics}.}

For example, if we wished to use physical fitness in scholarship deliberations, we would begin by proposing a causal narrative: how a student's character could lead them to a desire to accomplish difficult tasks through persistence; how this desire, in the right contexts, could cause them to train for a competitive sport; and how this training would cause them to improve in quantitative measures of physical fitness. We would be alert for signs that our reasoning was at fault---if excellence at handball is no more or less a signal of grit than excellence at squash, we should not prefer the squash player from Manhattan to the handball player from the Bronx.

This kind of reasoning, implicit or explicit, is found almost everywhere people gather to make decisions that affect the lives of others. Human-readable accounts of why someone failed to meet the bar for a scholarship, triggered a stop and frisk, or was awarded a government contract are the bread and butter for ethical debates on policy. These usually demand we explain both the role different perceptions played in the decision-making process (\emph{i.e.}, \emph{on what basis} the committee made the decision it did) and the causal origin of the facts that led to those perceptions (\emph{i.e.}, \emph{how it came about} that the person had the qualities that formed the basis of that decision).

Even if we agreed on a mechanism connecting a student's character and their physical fitness, we might be concerned with the causal role played by, say, socioeconomic status: a student's character may lead them to the desire to accomplish difficult tasks, but their socioeconomic status may rule out access to playgrounds and coaching. Should we agree on this new causal pathway, it might lead us to argue against the use of physical fitness, or to use it to make decisions only within a particular socioeconomic category.

\section{Clash of the Machines}

The demand for a causal account of both the origin of relevant facts and their use is at the core of the conflict between ethical decision-making and the use of big data. This is because the algorithms that use these data do not make explicit reference to causal mechanisms. Instead, they gain their power from the discovery of unexpected patterns, found by combining coarse-graining properties in often uninterpretable ways.

``Why the computer said what it did''---why one candidate was rated higher than another, for example---is no longer clear. On the one side, we have the inputs: data on each candidate. On the other side, we have the output: a binary classification (good risk or bad risk), or perhaps a number (say, the probability of a mortgage default). No human, however, wires together the logic of the intermediate step; no human dictates how facts about each candidate are combined together mathematically (``divide salary by debt,'' say) or logically (``if married and under twenty-five, increase risk ratio'').

The programmer sets the boundaries: what kinds of wirings are possible. But they allow the machine to find, within this (usually immeasurably large) space, a method of combination that performs particularly well. The method itself is usually impossible to read, let alone interpret. When we do attempt to represent it in human-readable form, the best we get is a kind of spaghetti code that subjects the information to multiple parallel transformations or unintuitive recombinations, and even allows rules to vote against each other or gang up in pairs against a third.

Meanwhile, advances in machine learning generally amount to discovering particularly fertile ways to constrain the space of rules the machine has to search, or in finding new and faster methods for searching it. They often take the form of black magic: heuristics and rules of thumb that we stumble on, and that have unexpectedly good performance for reasons we struggle to explain at any level of rigor. As heuristics are stacked on top of heuristics, the impact of these advances is to make the rules more tangled and harder to interpret than before. (This poses problems beyond the ethics of decision-making; the ability of high-powered machines to achieve increased accuracy at the cost of intelligibility also threatens certain avenues of progress in science more generally~\cite{krakauer2010intelligent}.)

As described by Ref.~\cite{alaimo}, the situation for the ethicist is further complicated because the volume of data demands that information be stripped of context and revealing ambiguity. Quite literally, one's behavior is no longer a signal of the content of one's character. Even if we could follow the rules of an algorithm, interpretation of the underlying and implicit theory that the algorithm holds about the world becomes impossible.

Because of the problem of euphemism, eliminating protected categories from our input data cannot solve the problem of interpretability. It is also true that knowledge of protected categories is not in itself ethically problematic and may in some cases be needed. It may aid us not only in the pursuit of greater fairness but also in the pursuit of other, unrelated goals.

The diagnosis of diseases with differing prevalence in different groups is a simple example. Consider the organ transplant case, and a protected category $\{a,b\}$. Individuals of type $a$ may be subject to one kind of complication, individuals of type $b$ equally subject to a different kind. Given imprecise testing, knowledge of an individual's type may help in determining who are the best candidates from each group, improving survival without implicit reliance on an ethically problematic mechanism. 

In other cases, fairness in decision-making might suggest we take into account the difficulties a candidate has faced. Consider the awarding of state scholarships to a summer camp. Of two candidates with equal achievements, we may wish to prefer the student who was likely to have suffered racial discrimination in receiving early opportunities. To do this rebalancing, we must, of course, come to learn the candidate's race.

Even when fairness is not an issue, knowledge of protected categories may aid decision makers well beyond the medical case described above. An undergraduate admissions committee for an engineering school might wish to rank a high-performing female applicant above a similarly qualified male, not out of a desire to redress wrongs or achieve a demographically balanced group, but simply because her achievements may be the more impressive for having been accomplished in the face of overt discrimination or stereotype threat~\cite{spencer1999stereotype}. A desire to select candidates on the basis of a universal characteristic (in this case, perhaps grit; see discussion above) is aided by the use of protected information.

In the organ transplant case, the knowledge of correlations may be sufficient. I need not know why group a and group b have the difference they do---only that I can gain predictive knowledge of their risk profiles by distinct methods. In the other two cases, however, discussions about how, when, and why to draw back the veil of ignorance~\cite{rawls2009theory} lead us to conversations about the causes and mechanisms that underlie inequities and advantages.

In sum, it is not just that computers are unable to judge the ethics of their decision-making. It is that the way these algorithms work precludes their analysis in the very human ways we have learned to judge and reason about what is, and is not, just, reasonable, and good.

\section{Algorithmic Solutions}

We find ourselves in a quandary. We can reject the modern turn to data science, and go back to an era when statistical analysis relied on explicit causal models that could naturally be examined from an ethical and political standpoint. In doing so, we lose out on many of the potential benefits that these new algorithms promise: more efficient use of resources, better aid to our fellow citizens, and new opportunities for human flourishing. Or, we can reject this earlier squaring of moral and technocratic goals, accept that machine-aided decision-making will lead to discrimination, and enter a new era of euphemisms, playing a game, at best, of catch as catch can, and banning the use of these methods when problems become apparent. Neither seems acceptable, although ethical intuitions urge that we trade utility (the unrestricted use of machine-aided inference) for more sacred values such as equity (concerns with the dangers of euphemism).

Some progress has been made in resolving this conflict. Researchers have begun to develop causal accounts of how algorithms use input data to classify and predict. The {\it contribution propagation} method introduced by Ref.~\cite{landecker2013interpreting}, or the {\it Bayesian list machines} of Ref.~\cite{letham2013interpretable} are two recent examples, that allow us to see how the different parts of an algorithm's input are combined to produce the final prediction. 

Contribution propagation is most naturally applied to the layered processing that happens in so-called deep learning systems.\footnote{Many machine-learning algorithms are designed to classify incoming data: mortgage applicants, for example, by degree of risk. They work by finding patterns in past data, and using the relative strengths of these patterns to classify new data of unknown type. ``Deep learning'' combines these two steps, simultaneously learning patterns and how they combine to produce the property of interest.} As a deep learning system passes data through a series of modules, culminating in a final prediction, contribution propagation allows us to track which features of the input data, in any particular instance, lead to the final classification. In the case of image classification, for example, it can highlight regions of a picture of a cat that led it to be classified as a cat (perhaps the ears, or forward-facing eyes); in this way, it allows a user to check to make sure the correct features of the image are being used. Contribution propagation can identify when an image is being classed as a cat solely on the basis of irrelevant contexts, such as the background featuring an armchair---in which cats are often found.

This makes contribution propagation a natural diagnostic for the ``on what basis'' problem: which features were used in the decision-making process. Applied to a complex medical or social decision-making problem, it could highlight the relevant categories, and whether they made a positive or negative contribution to the final choice.

The Bayesian list machine takes a different approach, and can be applied to algorithms that rely on so-called decision trees. A decision tree is a series of questions about properties of the input data (``is the subject over the age of twenty one?''; ``does the subject live on the south side?'') that result in a classification (``the subject is high risk''). Because the trees are so complex, however, with sub-trees of redundant questions, it can be hard to interpret the model of the world that the algorithm is using. The Bayesian list machine circumvents this problem by ``pruning'' trees to find simple decision rules that are human readable.

Neither advance can solve the full problem: simply knowing \emph{how} the variables were combined does not provide an explanation for \emph{why} the variables were combined in that fashion. While both methods allow us to look inside previously opaque black boxes, they leave us uncertain of the causal mechanisms that led to this or that combination of variables being a good predictor of what we wish to know. We may know ``on what basis'' the decision was made, but not how it came about that the individuals in question met the criteria.

Technical advances may go a long way to solving this second, more pressing problem. A long tradition in artificial intelligence research relied on building explicit models of causal interactions~\cite{pearl1997probabilistic}. Frameworks such as Pearl causality~\cite{pearl2000causality} attempt to create, in a computer, a mental model of the causes in the world expressed graphically, as a network of influences. This is, in spirit, similar to an earlier attack on the artificial intelligence problem---one described as ``good old fashioned'' artificial intelligence, or GOFAI, by John Haugeland~\cite{haugeland1989artificial}. GOFAI approaches try to make intelligent machines by mimicking a particular style of thinking---the representation of thoughts in a mathematical syntax, and their manipulation according to a fixed and internally consistent set of rules. The causal networks of Judea Pearl can be ``read'' by a human, and in a logically consistent fashion, their causal language can be used in moral explanations.

Pearl causality has had widespread influence in the sciences. But it does not yet play a role in many of the machine-learning algorithms, such as deep learning or random forests, in widespread use for monitoring and prediction today. One day, a framework such as this could provide a ``moral schematic'' for new algorithms, making it possible for researchers, policy makers, and citizens to reason ethically about their use. That day has not yet arrived.

\section{Information Theory and Public Policy}
\label{itpp}

In the absence of causal models that allow us to discuss the moral weight of group-dependent outcomes, progress is still possible. In this section, we will show how to encode a simpler goal: that decisions made by decision makers do not correlate with protected variables at all. One might informally describe such a system as ``outcome equal.'' Correctly implemented, our solution completely de-correlates category and outcome. Imagine you have heard that a person received a scholarship; using the outcome equal solution we present here, this fact would give you no knowledge about the race (or sex, or class, as desired) about the individual in question.

Whether such a goal is desirable or not in any case is, and has been, constantly up for debate, and we will return to this in our conclusions. The existence of a unique mathematical solution to this goal is not only of intrinsic interest. It also provides an explicit example of how technical and ethical issues intertwine in the algorithmic era. The exact structure of the mathematical argument has moral implications.

The method we propose post-processes and ``cleans'' prediction outputs so that we eliminate the possibility that the output of an algorithm correlates with a protected category. At the same time it will preserve, as much as possible, the algorithm's predictive power. The method alters the {\it outputs} of an algorithm, in contrast to recent work~\cite{friedler2014certifying} that has considered the possibility of altering an algorithm's {\it input}. Our recommendations here are also distinct from those considered by Ref.~\cite{starr2014evidence}, which seeks to exclude some variables as inputs altogether. Indeed, here, in order to correct for the effects of correlations, our calculations require knowledge of protected categories to proceed

To determine how to do this correction, this we turn to Information Theory. We want to predict a particular policy-relevant variable $S$ (say, the odds of a patient surviving a medical procedure, committing a crime, or graduating from college) and have at our disposal a list of properties, $V$. The list $V$ may be partitioned into two sub-lists, one of which, $U$, is unproblematic while the other, $W$, consists of protected variables such as race, sex, or national origin.

Given our discussion above, making a policy decision on the basis of $\mathrm{Pr}(S|V)$ may well be unacceptable. If it is unacceptable, so is using the restricted function $\mathrm{Pr}(S|U)$, because $U$ correlates with $V$ (the ``wrong side of the tracks'' problem). In addition, use of the restricted function throws away potentially innocuous use of protected categories.

We wish to find the distribution which avoids correlating with protected variables while minimizing the loss of predictive information this imposes. The insensitivity condition for this ``policy-valid'' probability, $\mathrm{Pr}_\mathrm{X}$ is 
\begin{equation}
\sum_{u} \mathrm{Pr}_\mathrm{X}(s,u,w) = \mathrm{Pr}(s)\mathrm{Pr}(w).
\label{ethics}
\end{equation}
Equivalently, $\mathrm{Pr}_\mathrm{X}(s|w)$---the probability of a protected category $w$ having outcome $s$---is independent of $w$, given the true distribution, $\mathrm{Pr}(w)$, of that category in the population. Our principle is thus that from knowledge of the outcome alone one can not infer protected properties of an individual. In the two examples above, allocation according to $\mathrm{Pr}_\mathrm{X}$ would mean that if you learn that a person received a life-saving transplant or was subject to additional police surveillance, you do not gain information about his race.

There are many $\mathrm{Pr}_\mathrm{X}$ that satisfy the constraint above. To minimize information loss, we impose the additional constraint that it minimize
\begin{equation}
\mathrm{KL}(\mathrm{Pr}_\mathrm{X}(S,V), \mathrm{Pr}(S,V)),
\end{equation}
where $\mathrm{KL}$ is the Kullback-Leibler divergence,
\begin{equation}
\mathrm{KL}(P,Q) \equiv \sum_{y} P(y) \log{\frac{P(y)}{Q(y)}}.
\label{kl}
\end{equation}
Minimizing Kullback-Leibler divergence means that decisions made on the basis of $\mathrm{Pr}_\mathrm{X}(S,V)$ will be maximally indistinguishable from the the full knowledge encapsulated in $\mathrm{Pr}(S,V)$ (the Chernoff--Stein Lemma; see Ref.~\cite{cover2006elements}).

Given the structure of Eq.~\ref{ethics} we can minimize Eq.~\ref{kl} using Lagrange multipliers. We require $|S||W|+1$ multipliers: one to enforce a normalization for $\mathrm{Pr}_\mathrm{X}$, and the remainder to enforce the distinct constraints implied by Eq.~\ref{ethics}. We find
\begin{equation}
\mathrm{Pr}_\mathrm{X}(s,u,w) = \mathrm{Pr}(s,u,w)\left[\frac{\mathrm{Pr}(s)}{\mathrm{Pr}(s|w)}\right].
\label{final}
\end{equation}
Knowledge of how predictions, $s$, correlate with protected variables $w$, allows us to undo those correlations when using these predictions for policy purposes.

Kullback-Leibler divergence has the property of becoming ill-defined when $\mathrm{Pr}(S,V)$ is equal to zero but $\mathrm{Pr}_\mathrm{X}(S,V)$ is not. However, the ethical intuitions that lead to the imposition of Eq.~\ref{ethics} do not apply when $\mathrm{Pr}(S,V)$ is precisely zero. This perfect knowledge case implies a very different epistemic structure: it is \emph{necessarily} true---as opposed to simply very probable---that a certain group can not have outcome $S$. Rather than the example of organ transplants or graduation rates, where such perfect knowledge is impossible, a better analogy is in the provision of pre-natal care. No notion of justice suggests that fair treatment requires equal resources to test both men and women for pregnancy. Correct accounting for these exceptions is easily accomplished, so that an agency can exclude men from pre-natal care, but, using the methods of this section, provide them optimally for women while preventing non-uniform allocation by race, religion, or social class.

The methods we present here are a mathematically clean solution to a particular problem. Use of these methods enforces a strict independence between protected categories and the variables one wishes to know.

In any specific case this may, or may not, be the desired outcome. Populations may be willing to trade off group-dependent outcomes in favor of other virtues. This can be seen in the ongoing debates over Proposition 209 in California (1996) and the Michigan Civil Rights Initiative (2006), both of which forbid the use of information concerning race to rebalance college admission, and the latter of which was affirmed as constitutional in April of 2014.  This ``outcome equal'' construction, in other words, does not absolve us of the duty to reason about the methods we use. Rather, it provides a limiting case against which we can compare other approaches. How much does a prediction change when we force it to be outcome equal? Do the changes we see alert us to potentially problematic sources of the correlations our algorithms use?

The relative transparency of decision-making in the era before data science allowed ordinary people to debate question of group-dependent outcomes on equal terms. Even when bureaucracies, traditions, or a failure to achieve consensus prevent them from implementing the changes they desire, citizens at least have had the ability to debate and discuss what they saw. When we use machines to infer and extrapolate from data, democratic debate on the underlying questions of fairness and opportunity becomes harder, if not impossible, for citizens, experts, and leaders alike. If we do not know the models of the world on which our algorithms rely, we cannot ask if their recommendations are just.

\section{Conclusions}

Machine learning gives us new abilities to predict---with remarkable accuracy, and well beyond the powers of the unaided human mind---some of the most critical features of our bodies, minds, and societies. The machines that implement these algorithms increasingly become extensions of our will~\cite{clark2008supersizing}, giving us the ability to infer the outcomes of thought experiments, fill in missing knowledge, and predict the future with an unexpected accuracy.

Over and above advances reported in the peer-reviewed literature, recent popular accounts provide a sense of the growing use of these tools beyond the academy~\cite{mayer2013big,siegel2013predictive}, and their use seems likely to accelerate in both scale (number of domains) and scope (range of problems within a domain).

Reliance on powerful but only partially-understood algorithms provides new challenges to risk management. For example, predictions may go wrong in ways we do not expect, making it harder to assess risk.  As discussed elsewhere~\cite{cate}, machine learning also provides new challenges to individual privacy: in a famous example, statisticians at WalMart were able to infer, from her shopping habits, a teenager's pregnancy before it became physically evident to her father~\cite{nyt}.

This chapter demonstrates the existence of a third challenge. This challenge persists even when algorithms function perfectly (when their predictions are correct, and their uncertainties are accurately estimated), when they are used by well-meaning individuals, and when their use is restricted to data and the prediction of variables, explicitly consented to by participants.

To help overcome this third challenge, we have presented the ambitious goal of reverse engineering algorithms to undercover their hidden causal models. We have also presented, by way of example, more modest methods that work in a restricted domain. In both cases, progress requires that we ``open the algorithmic box,'' and rely on commitments by corporations and governments to reveal important features of the algorithms they use~\cite{eden}. Judges already use predictive computer models in the sentencing phase of trials~\cite{starr2014evidence}. There appears to be no awareness of the dangers these models pose to just decision-making---despite the influence they have on life-changing decisions.

These are real challenges, but there is also reason for optimism. Mathematical innovation may provide the means to repair unexpected injustices. The same methods we use to study new tools for computer-aided prediction may change our views on rules we have used in the past. Perhaps most important, they may re-empower policy makers and ordinary citizens, and allow ethical debates to thrive in the era of the algorithm.

The very nature of big data blurs the boundary between inference to the best solution and ethical constraints on uses of that inference. Debates concerning equity, discrimination, and fairness, previously understood as the domain of legal philosophy and political theory, are now unavoidably tied to mathematical and computational questions. The arguments here suggest that ethical debates must be supplemented by an understanding of the mathematics of prediction. And they urge that data scientists and statisticians become increasingly familiar with the nature of ethical reasoning in the public sphere.

\vspace{0.5cm}
\noindent\emph{Acknowledgements}. I thank Cosma Shalizi (Carnegie Mellon University) for conversations and discussion, and joint work on the ``outcome equal'' solution presented above. I thank John Miller (Carnegie Mellon University), Chris Wood (Santa Fe Institute), Dave Baker (University of Michigan), Eden Medina (Indiana University), Bradi Heaberlin (Indiana University), and Kirstin G. G. Harriger (University of New Mexico) for additional discussion. This work was supported in part by a Santa Fe Institute Omidyar Postdoctoral Fellowship.

\clearpage


\end{document}